# Sensitive optical atomic magnetometer based on nonlinear magneto-optical rotation

Chris Hovde[*a], Brian Patton[b], Eric Corsini[b], James Higbie[c] and Dmitry Budker[b]
[a]Southwest Sciences Ohio Operations, 6837 Main Street, Cincinnati OH 45244;
[b]University of California, Dept. of Physics, 273 Birge Hall, Berkeley CA 94720;
[c]Bucknell University, Dept. of Physics and Astronomy, 174 Olin Science, Lewisburg, PA 17837

## ABSTRACT

A self-oscillating magnetometer based on nonlinear magneto-optical rotation using amplitude-modulated pump light and unmodulated probe light (AM-NMOR) in $^{87}$Rb has been constructed and tested towards a goal of airborne detection of magnetic anomalies. In AM-NMOR, stroboscopic optical pumping via amplitude modulation of the pump beam creates alignment of the ground electronic state of the rubidium atoms. The Larmor precession causes an ac rotation of the polarization of a separate probe beam; the polarization rotation frequency provides a measure of the magnetic field. An anti-relaxation coating on the walls of the atomic vapor cell results in a long lifetime of 56 ms for the alignment, which enables precise measurement of the precession frequency. Light is delivered to the magnetometer by polarization-maintaining optical fibers. Tests of the sensitivity include directly measuring the beat frequency between the magnetometer and a commercial instrument and measurements of Earth's field under magnetically quiet conditions, indicating a sensitivity of at least 5 pT/√Hz. Rotating the sensor indicates a heading error of less than 1 nT, limited in part by residual magnetism of the sensor.

**Keywords**: atomic magnetometer, heading error, diode laser, coherence

## 1. INTRODUCTION

Atomic magnetometers play an important role in applications including anti-submarine warfare, mineral exploration, and searches for buried objects. For typical air speeds, the magnetometer must be sensitive in the band from 0.01 Hz to 10 Hz. Atomic magnetometers usually measure total field (that is, the magnitude of the magnetic field vector). Total field measurements are not sensitive to the alignment of the magnetometer (except for small heading error effects), making deployment on a moving platform much simpler. In addition, atomic magnetometers feature outstanding accuracy and precision.

Research into atomic magnetometers has been revitalized by technical advances in atomic physics and optics, including techniques for producing coherence with long lifetime.[1,2] Sensitivity in the laboratory has approached and even exceeded that of super-conducting quantum interference devices (SQUIDs), without the need for cryogenics.[1,3] The availability of diode laser light sources delivered by polarization-maintaining optical fibers makes it possible to build instruments in which the magnetic measurement can be largely isolated from the magnetically noisy electronics needed to operate it.

Here we report an investigation of a magnetometer being developed for airborne magnetic anomaly detection. The technical approach is based on measuring the Larmor precession frequency of atomic alignment created and detected by nonlinear magneto-optical rotation using amplitude-modulated pump light and unmodulated probe light (AM-NMOR).[4,5] In this technique, a linearly-polarized, amplitude-modulated laser beam creates alignment in an atomic vapor. Alignment can be represented as coherence of the Zeeman sublevels of the atoms. Isotopically enriched rubidium, $^{87}$Rb was used in this work but other alkali atoms can also be used. The atomic vapor is held in an anti-relaxation coated glass cell, so that once created, the coherence persists for about 0.01 to 0.1 seconds. The coherence evolves in the presence of a magnetic field at the Larmor frequency (with a small correction for nonlinear Zeeman splitting). The evolution of the coherence periodically rotates the polarization of a linearly-polarized, cw read-out laser beam. The polarization rotation generally has signal components at both the Larmor frequency and twice the Larmor frequency.[6] The rotation of polarization is measured with a balanced polarimeter

---

set at 45° to the laser polarization, so that both outputs of the polarizer are of nearly equal intensity. The difference in the two outputs is detected. The prototype reported here has demonstrated that it can detect variations of about 5 pT in Earth's field of about 50 µT. Greater sensitivity is possible with improvements in the electronics and the atomic vapor cell.

## 2. METHODOLOGY

### 2.1 AM-NMOR Instrument

The prototype magnetometer based on AM-NMOR is outlined in Fig. 1. Linearly polarized, amplitude-modulated light from a diode laser is tuned to an atomic transition. Each pulse of light creates a coherence that is initially aligned along the light polarization axis. The aligned atoms then precess around the local magnetic field at a frequency (nearly) linearly proportional to the local magnetic field. The atoms rotate the polarization of the probe beam, and this is detected using a polarization analyzer. When the modulation frequency matches the precession frequency, the coherences created by each light pulse are in phase. The result is a resonant enhancement of the polarization rotation. The width of the resonance depends on the lifetime of the coherence. When the atomic vapor (isotopically enriched $^{87}$Rb, in this case) is sealed under vacuum in a specially prepared cell whose walls are coated with hydrocarbon wax, relaxation on the walls is suppressed. The magnetic spin relaxation lifetime in the cell used in this instrument is about 56 ms, which results in a sharp resonance width of about 2.6 nT with a peak amplitude of the polarization rotation of 1 to 10 mrad. The second factor that makes this approach so sensitive to magnetic fields is that the polarization rotation measurement can be made to high accuracy. Sensitivity can approach the shot noise level, permitting the center of the resonance to be identified to within 1 part in $10^5$. As a result, AM-NMOR can achieve sensitivity of better than 100 fT/√Hz.[7]

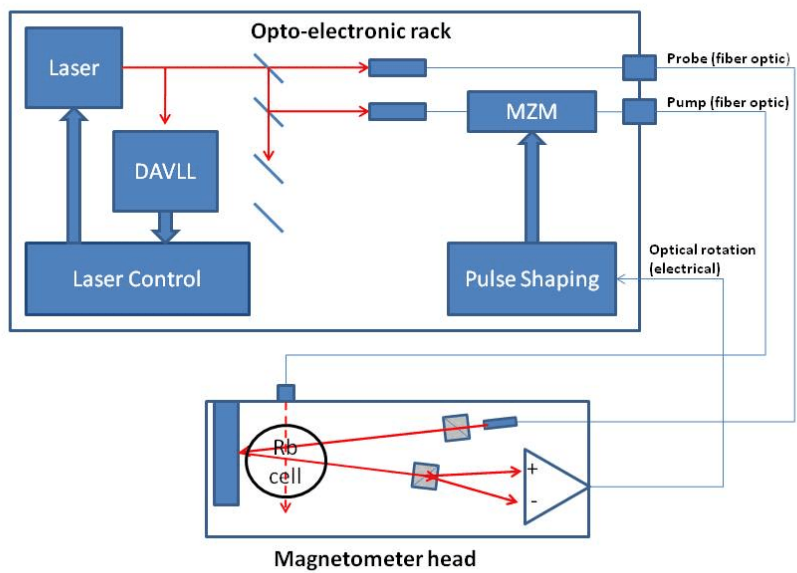

Figure 1 Schematic of the Phase II prototype, showing fiber-optic and electrical connections between the opto-electronics rack and one of two gradiometer channels. Dichroic atomic vapor line lock (DAVLL) used for stabilizing the wavelength of the laser. Mach Zender modulator (MZM) modulates the amplitude of the pump light. Pulse shaping electronics convert the analog optical rotation signal at twice the larmor frequency to a 20% duty factor pulse train at twice the Larmor frequency.

The operation of an AM-NMOR magnetometer requires careful control of the polarization and power of both the pump and probe beams as well as the vapor density within the cell. The power of the pump beam is adjusted to a point where the optical pumping rate is roughly equal to the relaxation rate of the coherence from other processes (wall collisions and spin-spin relaxation in the gas phase). An average power of about 100-200 µW is sufficient with the cells used. The probe power is adjusted to a slightly lower level, typically about 10-20 µW.

The light source is a distributed feedback diode laser (Eagleyard), which can access the F=1 and F=2 resonances of the $D_2$ transition in $^{87}$Rb near 780 nm. The laser temperature and injection current are stabilized with a commercial OEM controller (Wavelength Electronics). The laser beam is collimated, then split into several beams. One beam is split off to a compact dichroic atomic vapor line-lock (DAVLL) device[8,9] which is used to stabilize the laser wavelength at the point in the Rb spectrum which maximizes the magnetic sensitivity of the AM-NMOR signal. The DAVLL cell, including permanent magnets, is just 3.6 cm long.[10] Two beams (pump and probe) are coupled into polarization-maintaining optical fibers. The pump beam passes through a Mach Zender modulator (EOSpace, Inc.), operated to produce a ~20% duty cycle optical pulse. Pump and probe fibers are routed to the magnetometer head. The optical assembly includes two additional beams for a second magnetometer head.

The magnetometer head is built inside a G10 fiberglass tube of 17.8 cm in diameter by 50 cm long. The probe beam is nearly parallel to the axis of the G10 tube. It is launched from one end of the tube through a clean-up linear polarizer towards the atomic vapor cell mounted at the other end. The $^{87}$Rb atomic vapor is contained within a temperature-stabilized cell 5 cm in diameter by 5 cm long with a small stem containing a reservoir of $^{87}$Rb metal. A dielectric mirror reflects the probe beam back through the sample and through a Wollaston polarizer onto a pair of Si photodiodes. A differential trans-impedance amplifier (gain = 200 kΩ) is located adjacent to the photodiodes. The difference signal is carried back to the electronics area through twisted pair cable. The pump beam is launched from the side of the tube, perpendicular to the tube's axis, through the side of the cylindrical vapor cell. The pump polarization is also linear, with the electric field vector perpendicular to the tube axis (Fig. 2). With this arrangement of laser beams, the magnetometer is most sensitive when the magnetic field is parallel to the tube axis, in which case the polarization rotation signal is at twice the Larmor frequency.

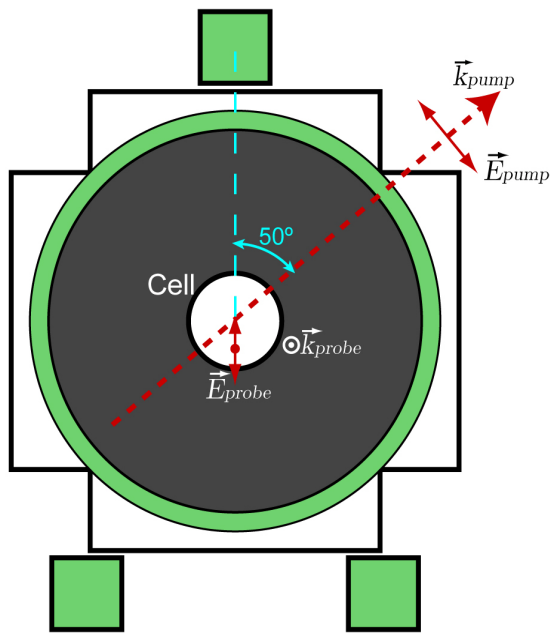

**Figure 2** End view of the magnetometer, showing the angles associated with the pump and probe beams.

The Rb cell heater is based on quad twisted heater cable from Lakeshore Cryogenics. Magnetic fields from the twisted pairs nearly cancel. To further reduce unwanted magnetic signals from the heater, it is energized by an ac supply operating at 1.7 kHz, so any residual magnetic field is ac and thus averages to zero over the time scale of the measurements for magnetic anomaly detection. The heater design was tested to make sure that the coolest part of the heated cell was the Rb reservoir in the side arm.

The magnetometer can be run in one of two modes. First, the magnetometer can be run as a forced oscillator, where a frequency generator is used to sweep the frequency of the laser amplitude modulator through the NMOR resonance. Typically, a lock-in amplifier is used to measure the amplitude of the optical rotation signal as a function of excitation frequency. The center of the resonance is used to determine the Larmor frequency and hence the magnetic field. The magnetometer can also be run in self-oscillating mode, where the optical rotation signal is amplified and fed back to the amplitude modulator. When the phase and gain of the feedback network are set correctly, the system spontaneously oscillates at the Larmor frequency or one of its harmonics. In this mode, a frequency counter is used to measure the oscillation frequency, or the oscillating signal can be digitized to post-analyze the frequency. Comparing the advantages of the two methods, self-oscillation provides rapid updates of the magnetic field, while forced oscillation is less susceptible to systematic errors resulting from changes in the phase of the signal.

For self-oscillation mode, a custom circuit board processes the optical rotation signal. This circuit includes an analog voltage amplifier (20k gain) that drives an analog filter (Allen Avionics, 10 kHz passband centered at 680 kHz; the pass band needs to be selected based on the local magnetic field), a Schmitt trigger and two monostables to create a phase-adjustable pulse with a 20% duty factor. Tests using a frequency-stabilized, low level reference signal as the input to the pre-amplifier indicate that the output showed jitter on the order of 20 mHz for 0.1 second counting intervals, equivalent to about 500 fT/√Hz. The pulses from this circuit board drive the Mach-Zender modulator. A dc level is also supplied to the modulator to optimize the contrast ratio. This dc level was adjusted using feedback from a circuit which measured the RMS fluctuations at an optical tap in the pump fiber.

## 2.2 Field site for measuring heading errors

The heading error tests were conducted in close coordination with Geometrics, which contracts to use the facilities at NASA Ames Research Center at Moffett Field to test their sensors for heading errors. The north end of the test building houses a magnetically-clean platform with three pairs of Helmholtz coils to compensate the ambient field if needed. The Helmholtz coils were not used in the experiment described herein, but additional coils in the anti-Helmholtz configuration were wound to null the ambient gradient at the location of the sensor.

The magnetometry test platform is located at the center of these coils and consists of a turntable mounted to a pivot, as shown in Fig. 3. The pivot axis is horizontal and perpendicular to Earth's field, and the entire mount can be rotated through nearly 120° dip angle in the (north/south)–(up/down) plane. This mount defines the plane of rotation of the turntable, which can be twisted as much as desired. The tube extending on the opposite side of the turntable from the magnetometer is an optical encoder which transmits information about the turntable angle to a computer. The encoder is weakly magnetic. The Phase II magnetometer was mounted to the turntable so that the geometric center of the atomic vapor cell was located at the intersection of the horizontal and vertical pivot points. In this way, rotation of the magnetometer was decoupled from translation through Earth's field. The centering accuracy is estimated at 1 mm. To the east and west of the pivot points, fixed witness sensors are mounted. These are Geometrics Cs magnetometers with an estimated precision of 5 pT.

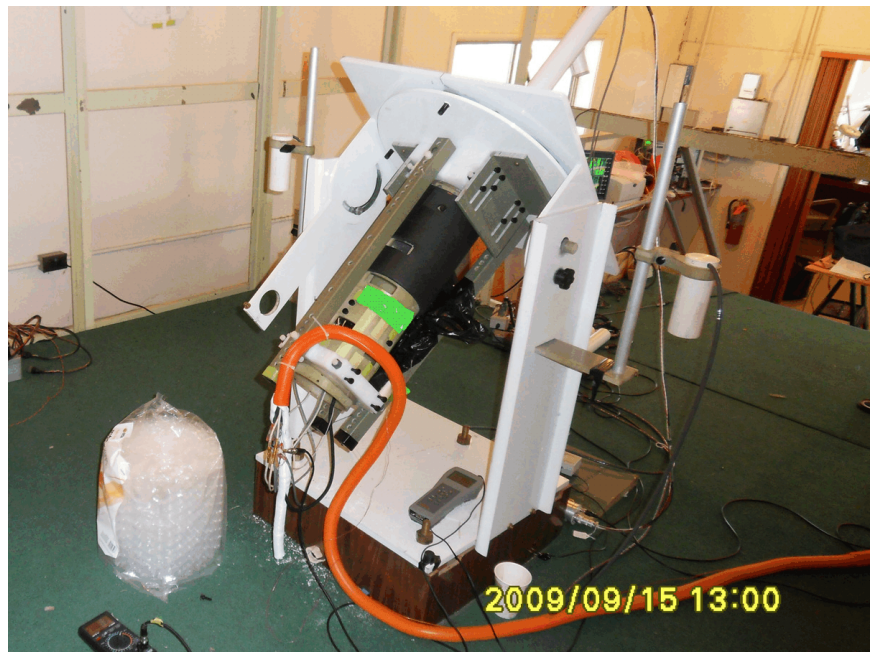

**Figure 3** The prototype magnetometer (gray and black tube with orange cable) mounted to the pivoting turntable (surrounding white structure). In the figure, the magnetometer is oriented along local Earth field. Two witness sensors (white cylinders) are mounted on either side of the pivoting turntable.

Tests were conducted on both the F=1 and F=2 resonances of $^{87}$Rb, as follows. First, the magnetic environment was mapped out using a commercial magnetometer, and the anti-Helmholz coils were adjusted to null the gradient at the test point. Next, the (nominally non-magnetic) metallic components of the magnetometer were carefully degaussed using a commercial video tape eraser placed next to the component, then slowly drawn away. The magnetometer was tuned up in self-oscillating mode by adjusting the cell temperature, pump and probe power, the laser wavelength, the phase delay of the pump pulse and its duration. The angle of the magnetometer in the vertical plane (dip angle) was adjusted to its desired value, then the turntable was twisted through 360° by hand, then returned to its starting value while recording data. The witness sensors are based on Cs, which has a Larmor frequency ½ that of $^{87}$Rb, and the NMOR sensor operates at twice the $^{87}$Rb Larmor frequency, so the NMOR sensor differed in operating frequency from the witness sensors by a factor of four. To facilitate comparison with the witness magnetometers, the $^{87}$Rb Larmor frequency of the magnetometer was divided by four using a 74ACT163 binary counter chip and counted by the same Geometrics counter that recorded the witness sensors, using a counting window of 0.1 to 1 second. (Separate tests of the divider showed it introduced jitter of about 4 mHz, so it did not contribute significantly to the error budget at the present level of sensitivity.) The divided AM-NMOR oscillation frequency nearly matched the Geometrics Cs magnetometer frequency except for small, nonlinear Zeeman terms and any heading error. Software written by Geometrics logged the output of the optical encoder, which encoded the twist angle, as well as the outputs of the two witness sensors and the Phase II magnetometer. The measurements were repeated at a series of dip angles.

## 2.3 Field tests to measure sensitivity

Tests of the sensitivity of the AM-NMOR magnetometer were conducted at the magnetic test facility at the Navy's Panama City support center. A two story wooden hut constructed with cedar wood and aluminum nails is located in a large, open wetlands. The hut is about 7 m by 7 m, set back from restricted-access roads by 40 m to the east and 60 m to the north. The hut is not heated or air conditioned. Weather during the week was very good, with no rainfall, light winds, and temperatures ranging from lows of about 50 °F to highs of up to 75°F. Electrical power is supplied by a single power cable running under a shaded structure from the north. The magnetometer head was installed on the upper floor of the hut. To maximize the distance from the magnetic measurement to the electronics, the electrical equipment was set up just outside the hut on its northern edge on a wooden and aluminum platform. A laser scalar gradiometer (LSG) based on four helium magnetometers (Polatomic) was set up in the hut to act as a witness sensor. The witness magnetometers have an effective noise floor of about 1.4 pT in the 0.1 to 10 Hz band, as judged from the spectral noise density (2pT/√Hz) of the difference between two sensors. A single witness sensor showed a noise density of about 5 pT/√Hz in the band from 0.1 to 10 Hz as well as narrow-band ambient noise at 60 Hz and its harmonics.

The noise measurements were made with a faster temporal resolution than that used in the heading error measurements. A lock-in amplifier was used to beat the NMOR signal against a nearby reference frequency down to an intermediate frequency of about 1 kHz, and this beat frequency was directly digitized at rates from 5 to 75 ksamples per second. Post-processing of the beat frequency, either by a local Fourier transform algorithm or by interpolating the zero crossing, yielded the frequency as a function of time.

## 3. RESULTS

### 3.1 Heading error measurements

The heading error results from both instrumental imperfections (such as magnetically permeable materials of construction, or currents in the electronics) and from changes in the AM-NMOR signal as a function of the changing angles between the laser beams and Earth's field (Fig. 2).

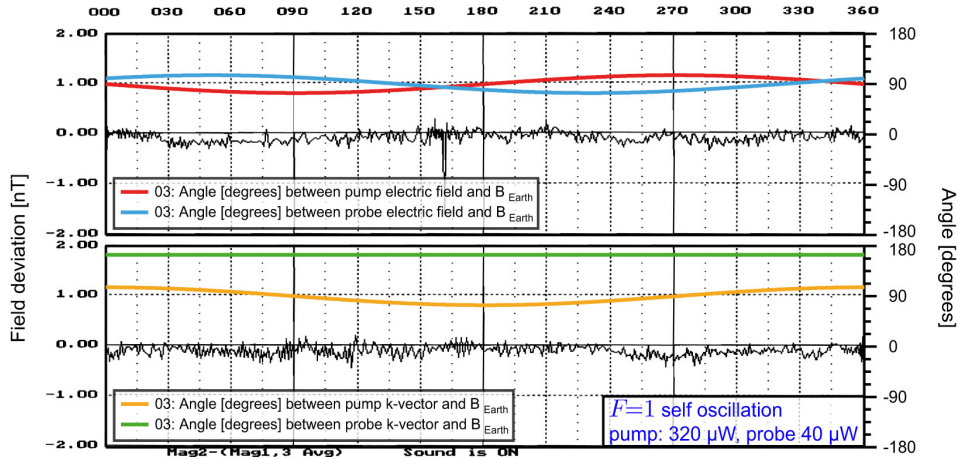

**Figure 4** Heading error observed for F=1 resonance in self-oscillating mode, when the k vector of the probe beam was aligned nearly parallel to Earth's field.

Figure 4 shows a typical heading error trace for the case where the sensor dip angle was chosen so that the magnetometer axis was aligned nearly parallel to Earth's field. In that orientation, the electric field vectors of the pump and probe beams are always close to perpendicular to the Earth field. The measured magnetic field relative to the field of the witness sensors is shown as the black trace. The top trace shows the measurements as the twist angle was increased from 0 to 360°, and the bottom trace shows the field as the twist angle was returned to 0°. The colored traces show the angles made by the electric fields of the pump and probe beams and by the k vectors of the pump and probe. For this case, the heading error is well under 1 nT.

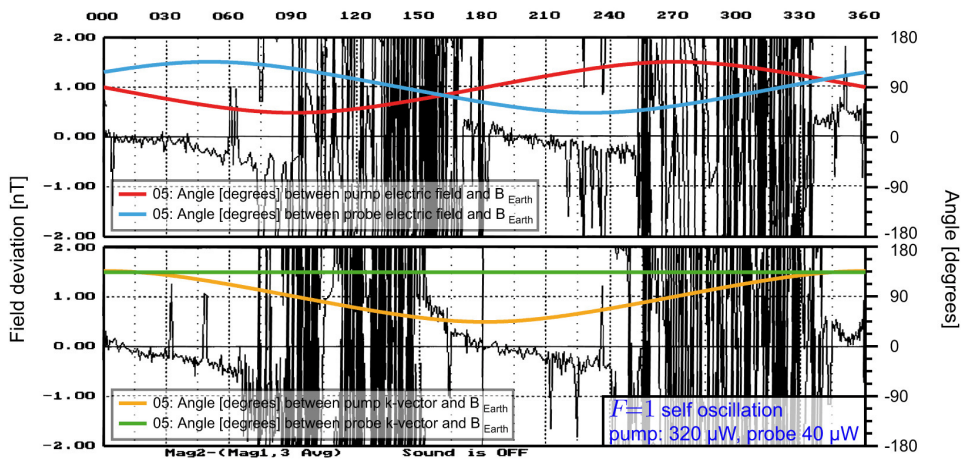

**Figure 5** Heading error measurements for F=1 self-oscillation, when the k-vector of the probe beam was at about 45 degrees to Earth's field. A dead zone is apparent for angles from 90° to 150° and 250° to 330°. Between these dead zones, the magnetic field varies by about 1 nT.

Increasing the dip angle between the magnetometer axis and Earth's field results in increased heading error. The results for F=1 are shown in Fig. 5. Two trends are evident. First, for twist angles roughly where the pump polarization is parallel to the Earth field, the output is extremely noisy. These regions are dead zones which result when the oscillator gain drops below unity, so the counting circuitry is counting noise. Between the dead zones, the oscillation frequency varies by about 1 nT. Similar measurements on the F=2 sublevel of $^{87}$Rb were obtained. The heading error approached ~3 nT for some of the F=2 measurements. The F=1 sublevel has only a single $\Delta m=2$ interval, whereas the F=2 sublevel includes three $\Delta m=2$ intervals split by the nonlinear Zeeman effect. A larger heading error is therefor expected for the F=2 sublevel.[12]

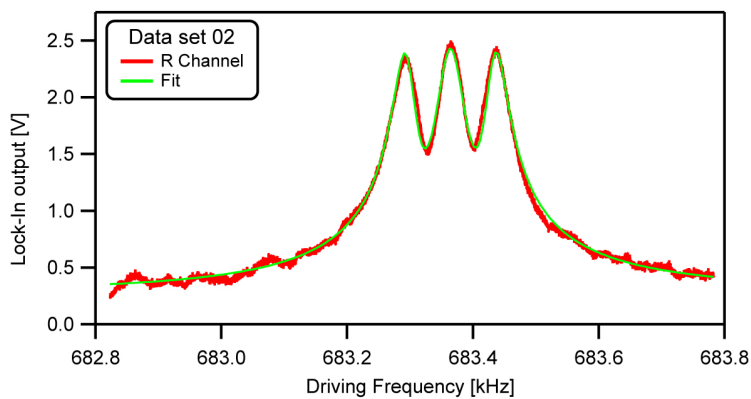

**Figure 6** Forced oscillation spectrum of F=2 resonance of $^{87}$Rb with pump power of 120 µW and probe power of 15µW.

The heading error was also studied by acquiring forced oscillation spectra (driving the modulator with a swept frequency source while performing lock-in detection of the polarimeter signal). To improve the signal:noise while minimizing the effects of drifts of the ambient field, 4 to 8 forced oscillation spectra were co-averaged while monitoring the witness frequency to make sure it didn't drift significantly. An example of such a spectrum is shown in Fig. 6. Fitting the magnitude of the spectrum (R= $\sqrt{(X^2+Y^2)}$) gives the center frequency of the pattern and the amplitudes and widths of the central and side peaks, without needing to define the lock-in phase. The statistical uncertainty in the center frequency, 42 mHz, is equivalent to an uncertainty in the magnetic field of 3 pT. The width associated with the spectrum of Fig. 6 is 54 Hz. At the laser powers used for the data shown in the figure, the ratio of the central peak height to the average of the heights of the two side peaks varies threefold as the angle between the magnetometer and the Earth's magnetic field is varied. Higher power gave an even wider range of ratios. The phase angle also varies, as is confirmed with a concurrent theoretical study at Berkeley.[11,12] As the magnetometer orientation is changed, the phase between the pump modulation and the observed probe rotation changes as well. To compensate for this additional phase delay, the self-oscillating magnetometer changes its frequency to eliminate any phase mismatch in the feedback loop. As a result, the oscillation frequency will shift as the heading changes.[4]

### 3.2    Testing the sensitivity

The magnetometer head was aligned so that the k vector of the probe laser lay along Earth's field. TheF=2 resonance was used. The self-oscillating signal was beat against a stable reference oscillator using a lock-in amplifier, and the low-pass filtered result was digitized and stored for later analysis. Data were recorded continuously for intervals ranging from thirty seconds to several minutes. The analysis included fitting subsets of about 0.01 seconds to a sine wave to determine the local beat frequency as a function of time, then determining the power spectrum from the time series of fitted frequencies. Results showed the same structure as observed in the witness sensors: noise decreasing with frequency from 0.01 Hz to a minimum at 10 Hz in the range 5 to 20 pT Hz$^{-1/2}$, then a strong 60 Hz peak with harmonics. The exception was two files recorded at the highest data rate, 75 kHz. These files showed a flat noise floor of about 100 pT. Figure 7 shows data from 5 Nov 2009, when one of the best noise floors was observed. In the 1 to 10 Hz band, the noise is about 4 pT/√Hz, similar to the ambient noise level (5 pT) observed from the witness sensors. These data were taken with pump power of 200 µW, probe power of 20 µW, and a sample rate of 5 kHz.

Other power spectra taken at faster digitization rates showed a clear peak associated with the 1.7 kHz heater current. At a frequency of 1 kHz, the minimum detectable signal was estimated to be 100 pT for a 1 second data acquisition time.

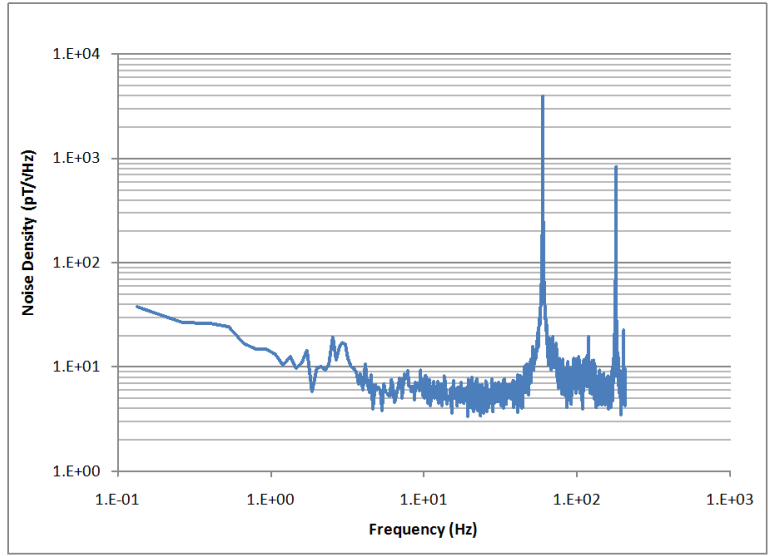

**Figure 7** Power spectral distribution from a time series recorded by the Phase II instrument. The noise floor near 20 Hz is about 4 pT/√Hz. Pump power is 200 µW, probe power is 20 µW.

The effect of optical power on the signal:noise of the magnetometer is shown in Fig. 8. The noise floor exhibits a clear decrease with pump power.

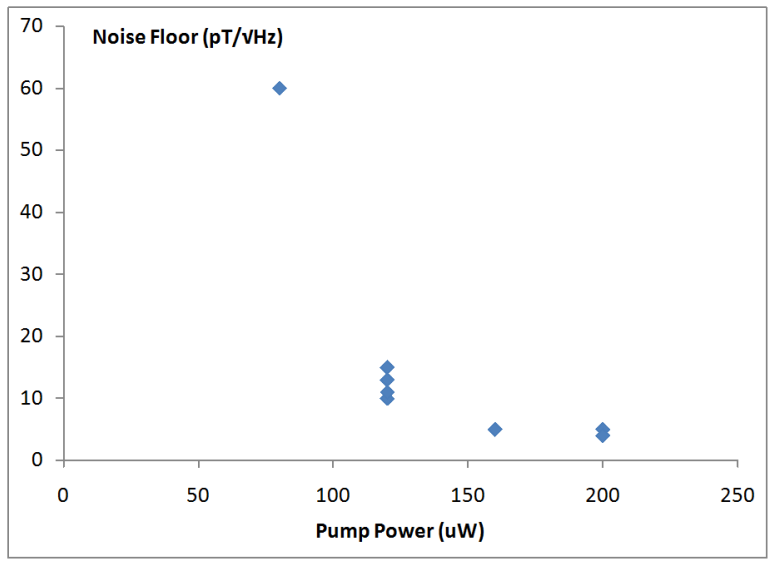

**Figure 8** Minimum noise around 10 Hz as a function of pump power.

## 4. CONCLUSIONS

A prototype AM-NMOR magnetometer has been built and field-tested at three sites, two of which are reported here. Significant strides were made in achieving stable performance, including heating the atomic vapor cell without introducing additional magnetic noise and locking the optical modulator. Improvements were made to the electronics that convert the polarimeter signal into pulses for the optical modulator. A compact DAVLL cell was incorporated into the optical head for stabilizing the wavelength of the laser.

The heading error was measured by comparing the self-oscillating frequency as the magnetometer was rotated to the field of two reference magnetometers that were fixed to the east and west. The heading error for the F=1 hyperfine component was less than 1 nT, while the heading error for the F=3 component was less than 3 nT. Forced oscillation spectra of the F=2 NMOR resonance showed that the ratio of the height of central resonance to the average of the heights of the outer resonances varied by a factor of three. These tests also mapped out the dead zone of the magnetometer. Related theoretical efforts at Berkeley have lead to a theoretical basis for the dead zone and heading error.

Measurements of the sensitivity of the magnetometer were performed at a magnetically quiet location at a Navy facility in Panama City, FL. The minimum sensor noise in the band from 0.1 to 10 Hz was 4 pT/√Hz, similar to the ambient magnetic noise (5 pT/√Hz) measured separately.

## ACKNOWLEDGMENTS

This research was supported in part by the Navy through contract number N68335-06-C-0042, by NASA through contract NNX07CA59P, and by the Dept. of Energy through grant DE-FG02-08ER84989.